\pdfoutput=1
\documentclass[11pt,a4paper]{article}

\usepackage[T1]{fontenc}
\usepackage[utf8]{inputenc}
\usepackage{lmodern}                  
\usepackage{mathptmx}                 
\usepackage[margin=1in]{geometry}
\usepackage{setspace}
\usepackage{microtype}
\usepackage{booktabs}
\usepackage{tabularx}
\usepackage[labelfont=bf,labelsep=space,skip=6pt]{caption}
\usepackage[hyphens]{url}
\usepackage[colorlinks=true,linkcolor=blue!50!black,citecolor=blue!50!black,
            urlcolor=blue!50!black]{hyperref}
\usepackage{xcolor}
\usepackage{apacite}                 

\AtBeginDocument{\renewcommand{\doi}[1]{\href{https://doi.org/#1}{\nolinkurl{https://doi.org/#1}}}}

\hypersetup{
  pdftitle={From the Gettier Machine to Justified True Output: A Three-Step Account of Rational Trust in Generative AI},
  pdfauthor={Tevfik Uyar},
  pdfkeywords={epistemology of artificial intelligence, justified true belief, Gettier problem, epistemic trust, large language models, reliabilism, critical thinking}
}

\newenvironment{criterion}
  {\par\medskip\begin{list}{}{\leftmargin=2.5em\rightmargin=2.5em}\item\relax}
  {\end{list}\medskip}

\title{\textbf{From the Gettier Machine to Justified True Output:\\
A Three-Step Account of Rational Trust in Generative AI}}
\author{Tevfik Uyar\\[2pt]
  \small ENTROPOL R\&D, Istanbul, Turkey\\
  \small \href{mailto:tuyar@entropol.com}{\texttt{tuyar@entropol.com}}\\
  \small ORCID: \href{https://orcid.org/0000-0003-0124-6910}{0000-0003-0124-6910}}
\date{}

\begin{document}
\maketitle

\begin{abstract}
\noindent
As generative artificial intelligence (AI) settles into everyday practices of inquiry, an old epistemological question returns in a new guise: when does a proposition a user takes from an AI output become something the user knows, and when is it rational to trust it as such? This paper argues that the characteristic failures of AI-mediated inquiry are precisely those that the justified-true-belief (JTB) tradition and its post-Gettier anti-luck refinements were built to diagnose, and it advances three claims. First, AI outputs inherit neither the presumptive warrant of testimony nor the reliabilist warrant of deterministic tools: unaudited trust in a calculator rests on a type-level reliability record, and for general-purpose generative models there is no type-stable process to which such a record could attach~--- a practical form of the generality problem. Second, use that stops at obtaining an output turns generative AI into a \emph{Gettier machine}: the user's fluency-based seeming of justification is disconnected from the truth of what is believed, and, as a clock running ten times too fast illustrates, sheer speed of production multiplies coincidental hits, inflating the appearance of reliability without improving the standing of any single hit. Third, the remedy is a three-step structure of use~--- acquisition, understanding, independent verification~--- whose first two steps answer to the classical belief and justification conditions and whose third answers to the post-Gettier sensitivity condition. The result is a stakes-indexed criterion of rational trust, \emph{justified true output} (JTO), with an everyday test: ``Had this been false, would I have noticed?'' The criterion is defended against objections from over-demandingness, the extended mind, model improvement, and redundancy.

\end{abstract}

\section{Introduction}\label{sec:intro}

An adult who double-checks a calculator's multiplication by hand strikes us not as a critical thinker but as somewhat obsessive. The same adult who copies a chatbot's five-paragraph historical narrative into a report without a single check strikes us as careless. The asymmetry between these two intuitions looks inconsistent at first sight, yet it tracks a sound epistemological distinction: technologies toward which trust is rational and technologies toward which trust has not~--- perhaps cannot~--- be rationally established do not share the same epistemic standing. The aim of this paper is to ground that distinction in the concepts of classical epistemology and to derive from it a normative framework for the rational use of generative AI outputs.

The practical problem is familiar. Generative AI tools produce fluent, coherent, self-assured text, and users routinely accept the propositions contained in that text without any check. The psychological mechanisms of this acceptance are comparatively well documented: fluently processed statements are judged more likely to be true \cite{reber1999}; humans instinctively posit minds behind language and humanlike cues, responding socially even to machines they know to be machines \cite{epley2007,nass2000}; and sustained reliance on automation breeds complacency and overtrust \cite{parasuraman1997}. The properly philosophical question, however, is not psychological but normative: under what conditions is a user's trust in a proposition contained in an AI output \emph{rational}, and under what conditions does that proposition attain the status of \emph{knowledge} for the user?

This paper's proposal is to answer with a three-step analysis of AI use. On this analysis, the cognitive relationship established with generative AI has three steps. At the first step the user employs the AI and obtains an output; the user has not reached the information~--- an instrument that reaches it has been employed. At the second step the user reads the output through, follows how its claims hang together, and reconstructs its reasons with their own conceptual resources. At the third step the user verifies the truth of the output's propositions against sources independent of the model. The central claim of the paper is that these steps are not an arbitrary usage recommendation but track the conceptual anatomy of knowledge itself. The first two correspond to the belief and justification conditions of the analysis that has been standard since Plato; the third corresponds to the anti-luck condition that the post-Gettier literature added to it~--- the condition that the truth of a belief not be accidental relative to the subject's grounds.

The normative upshot can be put as follows: \textbf{a proposition contained in an AI output attains, for the user, the status of \emph{justified true output} only when all three steps have been traversed, and only an output with that status rationally merits the kind of trust we extend to knowledge.} Use that stops at the first step is not knowledge even when the output happens to be true; it is at best accurate opinion, and its accuracy is owed to none of the user's epistemic achievements.

Three contributions are, I believe, distinctive. The first is a diagnosis of why AI outputs fall outside both the testimonial and the instrumental (``calculator'') classes of rational trust, which locates the decisive difference not in the model's indifference to truth as such but in the absence of a type-stable process to which a reliability record could attach (Section~\ref{sec:status}). The second is a relocation of the Gettier problem from the model to the user, together with an analysis of how speed of production inflates coincidental accuracy (Section~\ref{sec:gettier}). The third is a stakes-indexed criterion of rational trust and an everyday test that operationalizes the sensitivity condition at the moment of use (Section~\ref{sec:steps}).

The paper proceeds as follows. Section~\ref{sec:classical} briefly presents the Platonic analysis, the JTB tradition, and its anti-luck refinements; Section~\ref{sec:status} examines the epistemic status of AI output; Section~\ref{sec:steps} establishes the mapping between the three steps and the conditions on knowledge and states the criterion; Section~\ref{sec:gettier} analyzes the Gettier-type situations produced by step-skipping; Section~\ref{sec:objections} addresses four objections; Section~\ref{sec:applies} illustrates where the framework can be put to work; and Section~\ref{sec:conclusion} concludes.

\section{The Classical Analysis: Why True Opinion Is Not Knowledge}\label{sec:classical}

In the \emph{Theaetetus}, one of epistemology's founding texts, Socrates eliminates three successive answers to the question ``What is knowledge?'': knowledge is not perception; nor is it mere true opinion (\emph{orth\=e doxa}); the third candidate, examined at the dialogue's end, is that knowledge is ``true opinion accompanied by a \emph{logos}'' (\citeauthor{plato1997}, \emph{Theaetetus}, 201c--210b). The dialogue closes aporetically, without fully endorsing even this; yet the kernel of the definition~--- that a true belief must be accompanied by an \emph{account}, a \emph{reason}, to count as knowledge~--- became the main artery of Western epistemology. In the \emph{Meno} the binding role of reasons becomes still more concrete: the person who correctly guesses the road to Larissa gets the traveler there just as surely as the person who knows it; but true opinions, until tethered by ``reasoning about the cause'' (\emph{aitias logismos}), are like runaway slaves~--- they escape the mind (\citeauthor{plato1997}, \emph{Meno}, 97a--98a). What separates knowledge from opinion is not the accuracy of the outcome but the fact that the accuracy is an achievement creditable to the agent~--- a thought that contemporary virtue epistemology has made central \cite{greco2010}.

Twentieth-century analytic epistemology converted this analysis into the standard tripartite account: a subject S knows a proposition p only if (i)~S believes p, (ii)~p is true, and (iii)~S is justified in believing p. \citeauthor{gettier1963}'s \citeyear{gettier1963} celebrated counterexamples showed that the three conditions are not \emph{sufficient}: a subject can arrive at a justified true belief along paths in which justification and truth coincide only by accident, and in such cases our intuitions deny knowledge. The post-Gettier literature proposed further conditions to close this gap of ``epistemic luck'': that the belief be produced by a reliable cognitive process \cite{goldman1979}; that it be sensitive~--- had p been false, S would not have believed p \cite{nozick1981}; or that it be safe~--- S would not easily have been wrong in close possible circumstances \cite{sosa1999}. \citeA{pritchard2012} has argued that an adequate account needs both an anti-luck condition of this kind and an ability condition: knowledge is safe true belief that is creditable, to a significant degree, to the agent's cognitive abilities.

This paper does not aspire to settle the internal disputes of the JTB debate; it uses the framework as a \emph{diagnostic instrument}. For the troubles of AI-mediated inquiry are precisely the malfunctions this literature has spent decades mapping. The user who stops at the first step stands where the lucky guesser of the Larissa road stands; the user seduced by fluent prose mistakes the \emph{feeling} of justification for justification itself; and the cases in which a hallucination happens to be true are textbook-ready Gettier cases.\footnote{I frame the anti-luck condition in terms of Nozick's sensitivity rather than Sosa's safety because the question a user can put to herself at the moment of use~--- would I have noticed had this been false?~--- has the counterfactual shape of sensitivity. Nothing in the argument depends on this choice: the independent check that secures sensitivity in the cases at issue also makes it the case that S would not easily have believed p falsely, and so secures safety as well. The well-known difficulties for sensitivity as a necessary condition on knowledge (for instance, its conflict with closure) do not affect its use here as a criterion of rational trust in a specific class of outputs.} That a set of concepts assembled millennia and decades ago describes the new technology's epistemic predicament with such precision is itself evidence that the problem is not technological but perennial.

\section{The Epistemic Status of AI Output}\label{sec:status}

Before developing the three-step analysis, its object must be fixed: what, epistemically, \emph{is} a generative AI output? I will argue that three familiar models~--- the knowledge bank, the witness, the calculating device~--- are all wrong.

\subsection{Not a Knowledge Bank}\label{sec:bank}

The most widespread lay model casts the system as a vast encyclopedia: somewhere true information is stored, and the model fetches it on demand. But large language models do not \emph{fetch} information; they \emph{regenerate} text probabilistically on each occasion. What we call ``hallucination''~--- fabricated citations and facts~--- is therefore not a malfunction but a consequence of the system's normal mode of operation: the model draws no distinction between the process by which it produces true statements and the process by which it produces false ones. \citeA{hicks2024} argue on these grounds that model outputs count as \emph{bullshit} in \citeauthor{frankfurt2005}'s \citeyear{frankfurt2005} technical sense~--- speech produced with indifference to truth. The model does not lie, for lying requires knowing the truth and concealing it; the model is \emph{indifferent} to truth. The epistemological consequence must be stated with care. A reliabilist will reply that thermometers and calculators do not aim at truth either, and that what confers justification is a process's truth ratio, not its aim. The significance of truth-indifference here is narrower: because one and the same generative process yields true and false outputs alike, nothing in the output itself~--- its fluency, its confidence, its apparent reasons~--- discriminates the one from the other. Truth-indifference matters not because aiming at truth is a condition of justification, but because it removes any internal indicator by which a user could tell, from the output alone, which case she is in. Whether the process could nonetheless confer reliabilist justification is the question of Section~\ref{sec:calculator}.

\subsection{Not a Witness}\label{sec:witness}

A second model treats AI output as testimony. In the epistemology of testimony, the anti-reductionist tradition holds that trusting testimony carries a \emph{presumptive right}: absent counter-indications, accepting what one is told is rational by default (\citeNP{burge1993}; cf.~\citeNP{coady1992}). If that presumption transferred to AI output, stopping at the first step would be epistemically excusable. But the presumption's foundations are missing in the model. Testimonial warrant rests on the speaker's being an \emph{asserter}: one who vouches for the truth of what is said, bears responsibility for error, and is culpable for deceit. The model asserts nothing; it performs no act governed by a norm of truth \cite{hicks2024}. Two replies are available. With \citeauthor{lackey2008}'s \citeyear{lackey2008} statement view, one might hold that testimonial knowledge does not require a believing speaker, only a reliable statement; and \citeA{freiman2020} argue that technological devices may assert in a restricted sense they call quasi-testimony. Neither reply restores the presumption. Lackey pairs the speaker condition with a hearer condition: the recipient needs at least some positive reasons for accepting what she is told~--- which is precisely what step-one use lacks. And quasi-testimony, by its proponents' own account, differs from human testimony in that machines err in ways unlike human error and lack the moral standing of human testifiers; the anti-reductionist's presumptive right rests on exactly the features that are missing. As \citeA{hardwig1985,hardwig1991} showed, the modern order of expert-dependent knowledge is built on an \emph{ethics of trust}: in trusting an expert we trust, at bottom, the expert's epistemic conscience~--- that the data were not doctored, that competence is real, that the possibility of error is honestly flagged. The model has no epistemic conscience; the chain of vouching snaps at the model. This is why ``I asked the AI and it said so'' can never attain the epistemic force of ``I asked an expert and she said so'': in the latter there is a subject who assumes responsibility; in the former the responsibility is left hanging~--- or rather, it remains entirely on the user.

\subsection{Not a Calculator Either}\label{sec:calculator}

The third and subtlest error is to place trust in AI in the same class as trust in a calculator. Our trust in the calculator is legitimate, and its legitimacy is reliabilist through and through: the device executes a deterministic, human-programmed algorithm; identical inputs yield identical outputs every time; and its type-level reliability has, over decades of mass use, been tested across what are by now trillions of operations. Using a calculator is thus a belief-forming process whose truth ratio converges, for all practical purposes, on~1; and by the core reliabilist principle~--- that a process derives its justificatory power from its truth-conduciveness \cite{goldman1979}~--- the beliefs it issues are justified by process reliability. That is why we call the person who spot-checks its output obsessive: the additional check is a cost with an epistemic return of approximately zero.

\citeA{uyar2024} has called the terminus of such accumulated trust \emph{technological omnipotence}: the perception of a device's power, within a limited set of operations, as absolute and beyond question. Even a one-dollar, no-brand calculator has earned a ``narrow'' technological omnipotence with respect to arithmetic; no one crosschecks it, and, as just shown, no one rationally should. Today's large language models have not yet earned even this narrow omnipotence in most domains. For deterministic technologies, the accumulation of trust is epistemically \emph{legitimate}: the accumulation is the reliability statistic of a fixed process. Generative models lack the fixity on which such accumulation could rest: what is programmed is not the executed procedure but the model's manner of learning; the same question may receive different answers on different occasions; and the truth ratio of outputs swings with domain, topic, even prompt phrasing. To transfer the trust accumulated by deterministic processes onto a stochastic one is, in reliabilist terms, a \emph{category mistake}: there is no single, type-stable ``process'' whose measured reliability could serve as one's reason. This is a sharp, practical instance of what \citeA{conee1998} call the generality problem for reliabilism: every belief-forming process token instantiates many process types of widely differing reliability, and the theory owes an account of which type is the relevant one. For the calculator the problem is idle, since every candidate type~--- this device, this model, electronic arithmetic as such~--- has a truth ratio near~1. For a general-purpose generative model the candidate types~--- ``LLM output,'' ``this model's output,'' ``this model's output on legal citations,'' ``this answer to this phrasing of the prompt''~--- come apart dramatically, and the user typically has no way of knowing which type her case falls under, let alone its reliability. Technological omnipotence extended to generative AI is the illegitimate inheritance of the deterministic age by the stochastic one.

None of this implies that opaque computational processes can never ground justified trust. Against the worry that \emph{essential epistemic opacity}~--- an agent's inability to survey all the epistemically relevant elements of a process \cite{humphreys2009}~--- precludes justification, \citeA{duran2018} defend \emph{computational reliabilism}: trust in the results of opaque computational processes can be justified by reliability indicators such as verification and validation, robustness analysis, a history of successful implementation, and expert knowledge. The present account agrees and can be read as its user-side counterpart. Where such indicators exist for a well-delimited task type, they do for the user what the calculator's track record does; where they do not~--- as for most everyday uses of general-purpose models~--- the work they would have done has to be done at the level of the individual output, and that work is what steps two and three supply.

\subsection{AI as an Epistemic Technology}\label{sec:epistech}

What, then, is an AI output? In \citeauthor{alvarado2023}'s \citeyear{alvarado2023} term, AI is an \emph{epistemic technology}: a device that carries out information-related operations~--- searching, compiling, transforming, summarizing~--- while the epistemic value of those operations remains, in the last instance, dependent on human oversight. The definition fits the framework to be proposed exactly: an epistemic technology does not \emph{carry} knowledge; it \emph{accelerates} certain phases of the process of coming to know. The epistemic burden of that process~--- understanding and verification~--- remains non-delegably with the user. That fluent language renders the burden invisible does not mean the burden has been lifted. Humans are cognitively primed to equate polished, confident, frictionless text with true text; and, facing for the first time something that has language without a mind in the human sense, our intuitions are unprepared for the dissociation (cf.~\citeNP{reber1999}; \shortciteNP{bender2021}; \shortciteNP{heersmink2024}). This is precisely why an explicit normative framework, one that marks where the burden lies, is needed.

\section{The Three Steps and the Conditions on Knowledge: Correspondence and Criterion}\label{sec:steps}

\subsection{The Epistemological Content of the Steps}\label{sec:content}

Let us now recast the three-step analysis in epistemological terms. Let S be a user and p a proposition contained in an AI output.

\textbf{Step one~--- acquisition (the belief condition).} S poses a question to the model and receives an output containing p. At this stage S has not reached knowledge; an instrument has been employed~--- and, as aforementioned, what the instrument itself ``reaches'' is not knowledge but probabilistically generated text. S's relation to p at this stage is, at best, that of having acquired a belief: S believes p because the model said so. Only the first of the JTB conditions~--- belief~--- is met. The truth condition is undetermined; the justification condition is unmet. S may well have the impression that p is well supported~--- the text reads as reasoned~--- but the only ground S can actually offer for p is ``the model said so,'' and this ground can draw neither on the presumption of testimony nor on process reliability. (What to make of the impression itself is taken up in Section~\ref{sec:gettier}.) The user who stops at step one stands epistemically level with the lucky guesser of the Larissa road: the outcome may be accurate, but the accuracy is creditable to no epistemic achievement of S's.

\textbf{Step two~--- understanding (the justification condition).} S reads the output through, traces how its claims support one another, reconstructs the reasons with S's own conceptual resources, and stops to inquire wherever understanding fails. This is the exact counterpart of Plato's \emph{logos} condition: the ``bond of reasoning about the cause'' is fastened to true opinion at this step. A subtlety deserves emphasis: what justifies is not that reasons \emph{appear in} the output but that S \emph{appropriates} them. The model can upholster any output with reason-shaped sentences; reason-shaped text produced by a truth-indifferent system justifies no one so long as it remains unread and un-understood. In internalist terms, justification rests on reasons the subject can access and own; step two is where that access and ownership occur. Beyond the individual case, \citeA{uyar2024} has described where the institutionalized skipping of this step leads, in a theocratic simile: delivering verdicts without disclosing reasons is the signature of gods who send holy books, and trusting black-box systems whose grounds cannot in principle be examined is, structurally, an act of faith rather than an epistemic attitude. Step two is the line drawn against that attitude: a verdict whose reasons you cannot access is one you can only accept~--- never know. The \emph{Meno}'s runaway-slave metaphor explains the difference from the other side as well: content acquired through step two is tethered in the mind~--- durable, transferable to new contexts; content acquired at step one sits at copy-paste distance and travels no farther.

\textbf{Step three~--- verification (the anti-luck condition).} S tests the truth of p against reliable sources independent of the model~--- primary literature, official data, field checking, a competent human expert. A conceptual clarification is needed here, and it qualifies the correspondence: no audit can \emph{guarantee} truth; the truth condition is discharged by the world, not by the subject's efforts. Step three therefore does not correspond to the truth condition in the way step two corresponds to the justification condition. Its epistemological function is not to manufacture truth but to render S's belief-forming process \emph{sensitive} to it: verification institutes a check such that, had p been false, S would very likely have caught it. This is the practical embodiment of \citeauthor{nozick1981}'s \citeyear{nozick1981} sensitivity condition~--- ``had p been false, S would not have believed p''~--- that is, of the anti-luck condition that the post-Gettier literature added to the classical analysis precisely to ensure that truth and justification do not coincide by accident (see Section~\ref{sec:gettier}). Step three is where the truth condition is \emph{audited}: not made true, but brought into counterfactual dependence on the facts.

\subsection{The Criterion: Justified True Output}\label{sec:criterion}

The analysis can be gathered into a criterion of rational trust in AI results:

\begin{criterion}
\textbf{(JTO) The Justified True Output Criterion.} Let p be a proposition contained in an AI output, and let S's context of use be one in which the falsity of p would matter and would not otherwise come to S's notice. In such a context, p becomes a candidate for knowledge for S, and S's extending to p the kind of trust extended to knowledge is rational, if and only if the following jointly hold:

\textbf{(S1) Acquisition:} S believes p by way of the output;

\textbf{(S2) Model-independent justification:} S has understood the reasons for p by examining the output, has reconstructed them with S's own conceptual resources, and has appropriated them~--- such that S can now defend p on those reasons rather than because ``the model said so'';

\textbf{(S3) Independent verification:} the truth of p has been tested by at least one process independent of the model and of known reliability~--- such that, had p been false, S would very likely have noticed through that process.
\end{criterion}

The echo in the name is deliberate: JTO is the classical JTB analysis translated into the practice of AI use~--- from justified true \emph{belief} to justified true \emph{output}. Three further points deserve emphasis. First, JTO is not a \emph{definition of knowledge} but a \emph{criterion of rational trust}: as the Gettier literature shows, the complete definition of knowledge remains contested; but however that dispute is resolved, it remains defensible that treating an output as knowledge before S2 and S3 have been traversed is irrational. The criterion is in this sense robust to the definitional controversy: it employs necessary conditions drawn from JTB and its anti-luck refinement and commits itself to nothing about their joint sufficiency for knowledge.

Second, S2 and S3 are mutually irreducible and cannot substitute for one another. S3 without S2~--- verification without understanding~--- is possible but fragile: a user can check discrete facts, yet cannot select \emph{which} claims bear the load without understanding the output. In \emph{lossy} operations such as summarization and translation the danger is not false information so much as missing information one does not know is missing: the model has decided what to discard, and the discarded sentence may be the most critical one for the user's purpose. Detecting an absence is not a matter of fact-checking but of understanding source and context~--- that is, of S2. S2 without S3~--- understanding without verifying~--- is defenseless against persuasive text on rotten foundations: an output of flawless internal coherence may rest on a fabricated citation. Internal coherence checking (S2) and external truth checking (S3) close the Gettier gap from two different sides. How much work each does varies with the content. For a simple factual proposition~--- a date, a figure, a name~--- S2 is correspondingly thin: it consists in understanding exactly what is being claimed and what role the claim plays in the output, and most of the work falls to S3. For an argument, a summary, or an interpretation, the balance reverses.

Third, JTO applies \emph{in degrees}, which is why its antecedent restricts it to contexts in which a false p would matter and would go unnoticed. Demanding full verification of every AI output would be as irrational as demanding a hand check of every calculator result (see Section~\ref{sec:overdemand} for the over-demandingness objection). The criterion's practical calibration tracks the \emph{epistemic stakes} of the context of use, and its everyday calibrating instrument is a single question: \textbf{``Had this been false, would I have noticed?''} If the answer is yes~--- the output lies in a domain the user can test independently, or its falsity would betray itself in use~--- the sensitivity condition is already implicitly satisfied and further verification is wasted cost. If the answer is no, the user is operating on an insensitive belief, and there are steps left to climb: read, understand, verify. This one question is the pocket translation of Nozick's modal condition, and it moves critical thinking from the realm of abstract virtue into the moment of decision.

\subsection{The Correspondence in Summary}\label{sec:summary}

Table~\ref{tab:correspondence} carries a historical implication worth pausing on: the three-step framework is not a new ethics invented for AI but the twenty-four-century-old conceptual anatomy of knowledge, together with its twentieth-century anti-luck refinement, translated into a user's manual. What AI has done is not to create a new epistemological problem but to industrialize an old one: for the first time, we face a technology that produces the \emph{appearance} of justification in unlimited quantity at zero cost.

\begin{table}[htbp]
\centering
\caption{The correspondence between the conditions on knowledge and the three steps of AI use}
\label{tab:correspondence}
\small
\renewcommand{\arraystretch}{1.25}
\begin{tabularx}{\textwidth}{@{}>{\raggedright\arraybackslash}p{0.22\textwidth}
                               >{\raggedright\arraybackslash}p{0.30\textwidth}
                               >{\raggedright\arraybackslash}X@{}}
\toprule
\textbf{Condition on knowledge} & \textbf{Three-step counterpart} & \textbf{Failure mode when absent}\\
\midrule
Belief (\emph{doxa}) & S1~--- Acquisition & (No epistemic relation at all)\\
Justification (\emph{logos}) & S2~--- Examination and understanding & Unowned, borrowed belief; the fluency illusion\\
Non-accidental truth (\emph{al\=etheia}; post-Gettier sensitivity) & S3~--- Independent verification & Accuracy left to luck; Gettier cases\\
\bottomrule
\end{tabularx}
\end{table}

\section{The Price of Skipping Steps: AI as a Gettier Machine}\label{sec:gettier}

The structure of \citeauthor{gettier1963}'s \citeyear{gettier1963} original cases is this: the subject's belief is both true and justified, yet its truth flows not from the fact the justification points to but from a coincidence of which the subject is unaware. Epistemology's classic illustration is Russell's stopped clock: a stopped clock shows the right time twice a day, and the subject who ``learns'' the time happens to look at just such a moment \cite<for its relation to the Gettier structure, see>{heathcote2012}.

A clarification is needed before the structure is applied to AI use, since Section~\ref{sec:content} denied that the step-one user is justified. Gettier cases require a belief that is justified at least by the subject's own lights, and the step-one user is in exactly that position: the output presents itself as well reasoned, and it \emph{seems} to the user that p is supported. On a liberal internalism such as phenomenal conservatism, on which seemings confer at least some prima facie justification \cite{huemer2007}, the step-one user has a justification of this thin, seeming-based kind~--- and it is precisely this justification that is disconnected from the truth of p. On a more demanding view, the step-one user has no justification at all and stands where the lucky guesser of the Larissa road stands. Either way the belief falls short of knowledge; the Gettier diagnosis captures the more insidious case, in which the user has every subjective reason to take herself to know.

The generative-AI age requires an update to the example, because the stopped clock is the generative model's precise opposite: it is motionless, it always says the same thing, and for just that reason it sooner or later betrays its own defect. Instead of stopping the clock, speed it up. Consider a wall clock running ten times too fast. On a twelve-hour dial the gap between display and real time widens by nine hours every hour, and each time the gap completes a full twelve hours~--- that is, every eighty minutes of real time~--- the clock once again shows exactly the right time. The stopped clock is right twice a day; the tenfold-fast clock is right \emph{eighteen times} a day. Now let the multiplier grow. In general, a clock running $k$ times too fast is right $2(k-1)$ times a day: a hundredfold-fast clock is right 198 times, a thousandfold-fast clock 1,998 times, and nothing stops the count from growing without bound. The result is paradoxical at first sight: the faster the clock is broken, the more often it is coincidentally correct. Yet the epistemic situation has not improved by a millimeter. Each passage through the correct time is shorter in exact proportion to the clock's speed, so the chance that a given glance catches a correct reading is the same for the stopped clock, the tenfold clock, and the thousandfold clock alike.\footnote{With a tolerance of one minute, the stopped clock reads correctly for two minutes a day. A clock running $k$ times too fast passes through the correct time $2(k-1)$ times a day, but each passage lasts only $1/(k-1)$ of a minute; the total is again two minutes a day, whatever the value of $k$.} What speed multiplies is the \emph{count} of hits; what it leaves untouched is the standing of any single reading. The frequency of accuracy can be inflated at will, by speed alone, without the process acquiring the slightest connection to truth.

The analogy has a limit that should be stated at once. The fast clock's hits are rare relative to its readings, whereas a large share of a model's outputs are true; its truth ratio is far higher than the clock's. The clock is offered not as a model of the truth ratio but of a mechanism: what the user observes is a growing count of hits, and that count is a function of the volume of production rather than of any connection between the user's grounds and the truth. A high truth ratio would matter if the user could know which type her case falls under (Section~\ref{sec:calculator}); without that, the remembered string of hits supports only the generic expectation that outputs are often right, not the belief that \emph{this} output, on \emph{this} topic, is.

Generative AI is this clock with the multiplier turned up as far as it will go~--- not stopped but running at a furious pace; it is, so to speak, a \emph{Gettier machine}. The model produces text through a truth-indifferent process at a speed incommensurable with human scales; a substantial portion of what it produces is in fact true, because the statistical regularities of training data converge, much of the time, on true statements. Speed is no innocent quantity here; it is the multiplier of the illusion. A system generating thousands of statements per second multiplies its coincidental hits at the same rate, and in the user's eyes every hit lays one more brick in the edifice of trust. The stopped clock, with two hits a day, could persuade no one; the thousandfold clock, right nearly two thousand times a day, begins to give the impression of a working instrument. This is exactly how the illusion of accuracy is produced: the \emph{frequency} of the hits rises while the \emph{nature} of the hits~--- luck~--- stays fixed. And the fast clock is more dangerous than the stopped one, provided we hold fixed how it is consulted: not watched, but glanced at, as a model is consulted by being prompted. The stopped clock's display is frozen, so anyone who glances twice exposes the defect; the fast clock shows a different, perfectly ordinary time at every glance~--- precisely like a model that answers every question with fluent, coherent, confident, and each time different text. No single reading betrays the defect; it shows only to someone who sets the readings against an independent timepiece. Every true belief acquired by a user who stops at step one is, structurally, a fast-clock case: the belief is true and seems reasonable to its holder, but the link between its truth and the user's ground is coincidental. The user's ``ground''~--- the persuasiveness of the text~--- is an indicator that fails to discriminate true outputs from false ones; the model serves falsehoods with the same fluency and the same confidence. In Nozick's idiom: had p been false, the user would have believed p all the same. Insensitive beliefs are not knowledge even when true.

The kinship between LLMs and the Gettier problem has begun to be noticed in the recent literature; \shortciteA{mugleston2025}, for instance, observe that even the models' correct outputs are tainted by epistemic luck, so that LLMs may be counted ``victims of the Gettier problem.'' The present analysis relocates the problem. That literature asks whether the \emph{model} knows; our question is whether the \emph{user} can come to know through the model. The structure the Gettier-machine metaphor picks out arises not in the model's internal states but in the epistemic position of the user who stops at step one~--- which is why the remedy lies not in model engineering but in the steps the user must climb.

The practical value of the analysis is that it disarms the defense ``but it is right most of the time.'' Of course it is right most of the time; as the fast clock shows, the count of coincidental hits can be inflated indefinitely without the slightest epistemic improvement in the process. This is precisely the lesson epistemology drew from Gettier: knowledge is a matter not of hit rate but of \emph{how} the hits are secured. Steps two and three are the mechanisms that move AI-mediated belief from the regime of luck to the regime of achievement: S2 converts the user's ground from ``the text was persuasive'' to ``I have reconstructed the reasons myself''; S3 renders the belief sensitive to truth. Together they transform the person glancing at the fast clock into a person whose clock runs true and has been set against an independent timepiece.

\section{Objections and Replies}\label{sec:objections}

\subsection{The Over-Demandingness Objection}\label{sec:overdemand}

\emph{Objection:} Applied consistently, JTO paralyzes everyday epistemic practice. No one hand-checks the calculator, no one verifies every encyclopedia entry against primary sources; the testimonial order of knowledge already runs, in large part, on unaudited trust \cite{hardwig1985}. Imposing a special burden of audit on AI is arbitrary discrimination.

\emph{Reply:} The objection misses the distinction on which the criterion rests. What the legitimate cases of unaudited trust share is that the trust is underwritten by a \emph{type-level record of reliability}: the calculator's deterministic process has been tested across trillions of operations; behind the encyclopedia stand editorial control, refereeing, and institutional reputation; behind the human witness operate the norms of responsibility and vouching that frame the act of assertion. In the generative model all three underwritings are absent: the process is stochastic and its reliability swings from domain to domain; the output has passed through no editorial control; the model bears no assertoric responsibility. JTO invents no new burden for AI; it locates the burden that other sources carry for us and that, in AI, is left unassigned~--- namely, on the user. Moreover, the criterion is applied in degrees from the outset (see Section~\ref{sec:criterion}): the test ``Had this been false, would I have noticed?'' exempts low-stakes and self-testing uses~--- feedback from code that runs, texts in the user's own domain of expertise~--- from the burden of audit. What is demanded is not the verification of every output but the refusal to transact high-stakes business on insensitive beliefs.

\subsection{The Extended-Mind Objection}\label{sec:extended}

\emph{Objection:} On \citeauthor{clark1998}'s \citeyear{clark1998} extended-mind thesis, cognitive processes are not bounded by the skull: Otto's notebook, under the right conditions, is part of his memory. If AI tools count as extensions of the user's cognitive system, there is no principled difference between ``the model says'' and ``my memory says''; what the analysis segregates as step one is simply the ordinary functioning of an extended mind.

\emph{Reply:} The extended-mind thesis states explicit conditions for an external device's inclusion in the cognitive system: the device must be reliably available; its deliverances must be automatically endorsed; and~--- the critical condition~--- its contents must be there \emph{because they were consciously endorsed in the past}. Every line in Otto's notebook was once written by Otto; the notebook is a depository of Otto's past epistemic labor. The generative model's output is the product of no past endorsement by the user; the content is produced at query time, independently of the user's labor. AI therefore resembles not Otto's notebook but a notebook of unknown authorship, pressed into Otto's hand, whose pages can change from time to time. The extended-mind framework thus supports JTO rather than undermining it: steps two and three are precisely the operations that pass content through the condition of conscious endorsement and render it \emph{includable} in the extended cognitive system. When the steps are traversed, AI becomes a genuine cognitive extension; when they are skipped, the mind has not extended~--- it has merely surrendered to incoming text, and the question then is who is using whom. A qualified parallel can be found within the extended-mind tradition itself: examining a particular retrieval-augmented configuration~--- a GPT model coupled to Andy Clark's own corpus (``Digital Andy'')~--- \citeA{smart2025} conclude that the system they study fails the criteria of active externalism, because the subject does not actively trigger the integration, while explicitly leaving open that more agentic, actively integrated architectures might yet satisfy them. Their diagnosis converges with the present one: what canonical extension requires, and what step-one use lacks, is precisely active engagement~--- which, on this framework, is exactly what steps two and three supply.

\subsection{The ``Models Are Improving'' Objection}\label{sec:improving}

\emph{Objection:} Hallucination rates are falling; models increasingly cite sources and report uncertainty. JTO elevates a passing teething problem into a permanent epistemological principle.

\emph{Reply:} Two answers. First, the criterion rests not on the current error rate but on the two features identified in Section~\ref{sec:status}: the absence of an internal indicator that discriminates true from false outputs, and the absence of a type-stable process with a known reliability record. A falling error rate changes neither: the model speaks truth more often while its outputs remain indistinguishable, from the inside, whether true or false. When the tenfold-fast clock is replaced by one that loses a second a day, the Gettier structure thins but does not vanish~--- what makes the user's belief sensitive to truth is still the audit outside the model. Second, the improvement itself can be measured only through the institutionalized form of step three~--- benchmarks, independent evaluation, expert oversight; even the proposition ``models are improving'' cannot be known without S3. If someday, for specific task classes, specific models compile a type-level reliability record approaching the calculator's, JTO will legitimately lighten the burden of audit for that class; the criterion is open to this, because it indexes the burden not to the technology but to the absence of proof of reliability. Until that point is declared, the omnipotence extended to general-purpose generative models is trust that precedes evidence~--- and trust that precedes evidence is, by the standards of epistemic rationality, premature. There is, moreover, a self-feeding dynamic here: as the attribution of infallibility grows, so does the authority delegated to the system; as delegated authority grows, the system's ``track record'' swells and reinforces the attribution \cite{uyar2024}. The steps on which JTO insists are the individual-level breaking point of that loop: the user who preserves the habit of verification refuses to add evidence-free links to the chain.

\subsection{The Redundancy Objection}\label{sec:redundancy}

\emph{Objection:} If S3 requires checking p against a model-independent source, then once the check is made, S's belief is justified by that source, not by the model. JTO thus makes generative AI epistemically idle: the knowledge comes from elsewhere, and ``justified true \emph{output}'' is a misnomer.

\emph{Reply:} The objection is right about where the justification comes from and wrong about what follows. \citeauthor{reichenbach1938}'s \citeyear{reichenbach1938} distinction between the context of discovery and the context of justification applies: a hypothesis does not owe its justification to the circumstances of its discovery, yet those circumstances are not thereby idle. Generative AI contributes chiefly to discovery~--- it proposes candidate answers, surfaces considerations, drafts structures, and points to where to look. Its epistemic value lies there, and it is considerable, because checking is frequently far cheaper than finding: verifying that a proposed citation exists and says what is claimed may take minutes, while locating the relevant source unaided may take hours. Nor is the model's contribution confined to discovery. At S2, the reasons the user appropriates are often reasons the output first articulated; once reconstructed and owned, they justify on their own merits~--- which is just what it is for a reason to be appropriated rather than borrowed. ``Justified true output'' names the status an output acquires \emph{for the user} once she has done her part, not a status the output had on arrival. The objection thus restates the paper's thesis rather than refuting it: the model is an epistemic technology that accelerates phases of coming to know (Section~\ref{sec:epistech}), not a source whose say-so justifies.

\section{Where the Framework Applies}\label{sec:applies}

JTO need not remain an abstract analysis; the steps themselves constitute a teachable, applicable, and auditable framework. Since the criterion's normative burden scales with the scope of the harm a false output would cause, the framework operates across a range that runs from individual learning to professional knowledge production and institutional decision-making. A few illustrations will suffice.

Perhaps the most immediate use of the framework is diagnostic: determining whether a process of learning is really one. A process that stops at step one profits no mind: multiplying with a calculator teaches no mathematics, and putting questions to an AI is no substitute for the act of inquiry; a product appears, but no learning occurs. JTO makes the difference visible, because learning is exactly what happens at S2~--- the reconstruction of reasons with one's own conceptual resources; the student who delegates S2 has delegated not the homework but the learning itself. A curricular principle follows from the diagnostic: AI-literacy education need not reduce to tool training but can place \emph{step awareness} at its core; the question a student can ask at every use~--- ``Which step am I on right now?''~--- exposes in a single stroke the difference between the student who has AI do the assignment and the student who learns with AI. If the capacity to evaluate information objectively regardless of its source~--- even a seemingly infallible system~--- is the population-scale first line of defense against the slide from attributed infallibility to unquestioned acceptance \cite{uyar2024}, the weight this one question carries becomes apparent. The framework also shows why prohibition fails: a ban merely drives use underground, where it is necessarily stepless.

A second use is to translate the steps into professional norms in contexts where outputs are passed on to others~--- where error multiplies at social rather than individual scale. Journalism is the sharpest case: the ordinary user who stops at step one forfeits their own learning; the journalist who stops there misinforms a public. For the journalist the three steps are not a preference but a professional obligation, and JTO's components translate directly into the profession's principles. AI is never a source but an assistant that leads to sources; ``I asked the AI and it said so'' cannot appear in the language of news because, as Section~\ref{sec:witness} showed, in that sentence the chain of vouching is severed; every citation the model offers must be checked for existence and content (S3). Responsibility cannot be delegated: the byline under the story belongs to the journalist, not the model, and when a false summary is published the error belongs to the publisher of the summary. And the profession's core work~--- asking, doubting, picking up the phone to ask ``is this true?''~--- cannot be handed to the model, for that phone call is step three itself. The same translation can be repeated for every profession that acts on AI-mediated reports and analyses~--- lawyers, physicians, engineers, managers: as the stakes grow, the steps harden from advice into obligation.

Finally, the framework serves the individual user as an instrument of self-audit in everyday decisions. The test proposed in Section~\ref{sec:criterion}~--- ``Had this been false, would I have noticed?''~--- is the practical way of determining, at the moment of decision, which use demands which step: if the answer is yes, the user already occupies a context that satisfies the sensitivity condition; if no, there are steps left to climb. Thus JTO returns from its origin as a philosophical analysis to the educator's curriculum, the profession's code, and the user's daily habit~--- one criterion, translated three ways.

\section{Conclusion}\label{sec:conclusion}

This paper has tried to formulate the conditions of rational trust in generative AI results by leaning on epistemology's oldest analysis. The conclusion can be summarized as follows. The three steps of the relationship established with an AI output~--- acquisition, understanding, verification~--- are the practical counterparts of the conditions on knowledge set by the classical analysis and its post-Gettier refinement: belief, justification, and truth that is not accidental relative to one's grounds. Use that stops at the first step produces, at best, true opinion, and owes even that to the accuracy of a clock running a thousand times too fast: frequent, fluent, plausible on every glance~--- and lucky. Facing a technology that can produce the appearance of justification without limit while remaining structurally indifferent to truth, the burden of justification and verification stays, non-delegably, with the user; the measure of shouldering that burden is JTO, and its everyday test is a single question: ``Had this been false, would I have noticed?''

Let us close with the framework's more general implication. The history of technology is a history of delegating epistemic labor to machines, and most of those delegations~--- of calculation, of record-keeping, of search~--- proved legitimate. What distinguishes generative AI is that, for the first time, not only labor but the \emph{appearance} of labor has become producible: text that looks like reasons, citations that look like sources, an assistant that writes as if it understood. In such an age Plato's criterion is not an antiquarian scruple but a practical survival tool. The difference between knowledge and lucky opinion remained, for twenty-four centuries, a distinction that mostly concerned philosophers; generative AI has planted it inside a decision every person makes every day. The cost of abandoning the steps at the individual level is the loss of learning and knowledge; their collective abandonment ends in the condition \citeA{uyar2024} describes as technocratic theocracy~--- a settlement in which decisions are surrendered to a system whose reasons go unexamined and unquestioned, and in which critical thinking and human agency erode. The way to remain the principal actor~--- to use the instrument rather than be used by it~--- is to redraw that ancient distinction at every output.

\section*{Statements and Declarations}

\textbf{Competing interests:} The author declares no competing interests.\\
\textbf{Funding:} No funding was received for conducting this study.\\
\textbf{Data availability:} Not applicable; no datasets were generated or analyzed.

\bibliographystyle{apacite}
\bibliography{references}

\end{document}